\documentclass[preprint,amsmath,amssymb,aps,nofootinbib,showpacs]{revtex4}
\usepackage{graphicx}
\usepackage{bm}
\usepackage{subfigure}
\usepackage{amsmath}

\newcommand{\pr}{Phys. Rev.\ }

\newcommand{\jpa}{J. Phys. A\ }
\newcommand{\jpb}{J. Phys. B\ }
\newcommand{\njp}{New J. Phys.\ }
\newcommand{\etal}{{\em et al. }}

\newcommand{\UQ}{School of Mathematics and Physics, University of Queensland, Brisbane, 
QLD 4072, Australia.}

\begin{document}

\title{An entangling atomtronic beamsplitter}

\author{C.~V. Chianca and M.~K. Olsen}
\affiliation{\UQ}
\date{\today}

\begin{abstract}

We propose and analyse the use of a three-well Bose-Hubbard model for the creation of two spatially separated entangled atomic samples. Our three wells are in a linear configuration, with all atoms initially in the middle well, which gives some spatial separation of the two end wells. The evolution from the initial quantum state allows for the development of entanglement between the atomic modes in the two end wells. Using inseparability criteria developed by Hillery and Zubairy, we show how the detected entanglement is time dependent, oscillating with the well occupations. We suggest and analyse a method for preserving the entanglement by turning off the different interactions when it reaches its first maximum. We analyse the system with both Fock and coherent initial states, showing that the violations of the Hillery-Zubairy inequality exist only for initial Fock states and that the collisional nonlinearity degrades them. This system is an early step towards producing entangled atomic samples that can be spatially separated and thus close the locality loophole in tests of non-local quantum correlations.

\end{abstract}


\maketitle


In this letter we combine the two fields of quantum information and bosonic atomtronics to propose a method for the fabrication of spatially isolated entangled atomic populations. Bosonic atomtronics is field of investigation in which analogues of electronic circuits and devices are constructed using ultra-cold bosonic atoms rather than electrons as in conventional electronics~\cite{overview}. The conventional way to construct an atomtronic device is to use cold atoms trapped in an optical lattice, which has a description in terms of the Bose-Hubbard model for bosonic atoms~\cite{BHmodel}. Shortly after the realisation of trapped Bose-Einstein condensates (BEC), Jaksch \etal\cite{Jaksch} showed that this model can provide an accurate description of bosonic atoms trapped in a deep optical lattice. In this work we use a three well Bose-Hubbard model to propose and analyse a quantum atomtronic beamsplitter, demonstrating that this can split an initial condensate in the central well into two separated condensates which are entangled with each other. We then show how this entanglement may be preserved and make a suggestion for further spatial separation of the two parts of the bipartite entangled state.

Continuous-variable entanglement is an area of active research~\cite{Braunstein,Stefano}, with many of the obtained results only applying fully to Gaussian systems and measurements. Quantum information theory includes the study of entanglement between continuous-variable phase quadratures of bosonic fields, with a number of inequalities having been developed to detect the existence of the property. The most commonly used are those developed by Duan \etal\cite{Duan} and Simon~\cite{Simon}, using combinations of quadrature variances. More recently, Teh and Reid have shown the degree of violation of these inequalities that is necessary to demonstrate not just inseparability, but genuine entanglement~\cite{Teh}, as these are not necessarily the same for mixed states.
The criteria we will use here were developed by Hillery and Zubairy~\cite{HZ} and expanded on by Cavalcanti \etal\cite{ericsteer} to cover multipartite entanglement, steering, and violations of Bell inequalities. As shown by He \etal\cite{He}, the Hillery and Zubairy criteria are well suited to number conserving processes such as that of interest in this letter, but as shown by Olsen~\cite{toberejected}, they will sometimes miss actually existing entanglement in quantum optical systems.

Entanglement in condensed atomic systems has been predicted and examined in the processes of molecular dissociation~\cite{KVK}, four-wave mixing in an optical lattice~\cite{Campbell,Mavis,Andy4}, and in the Bose-Hubbard model~\cite{Hines}. In the latter it is produced by the tunnelling between wells, in both the 
continuous~\cite{Oberthaler2008,Oberthaler2011,He} and pulsed tunnelling configurations~\cite{myJPB,myJOSAB}. The quantum correlations necessary to detect entanglement can in principle be measured using the interaction with light~\cite{homoJoel,SimonHaine}, or by homodyning with other 
atomic modes~\cite{Andyhomo}. 
We note here that the entanglement we are examining is a collective property between atomic modes which are spatially separated, and is not between individual atoms~\cite{Mavis}. This point, unavoidable for indistinguishable bosons, has previously been raised by Chianca and Olsen~\cite{SU2Cinthya}, and was recently put on a formal basis, using the language of quantum information theory, by Killoran \etal~\cite{superobvious}.


In this letter we will follow the approach taken by Milburn \etal\cite{BHJoel}, generalisng this to three wells~\cite{Nemoto,Chiancathermal}, and using the fully quantum positive-P phase space representation~\cite{Pplus} rather than a three-mode Gross-Pitaevskii approach. We consider this to be the most suitable approach here because it is exact, allows for an easy representation of mesoscopic numbers of atoms, can be used to calculate quantum correlations, and can simulate different quantum initial states~\cite{states}. Just as importantly, the positive-P calculations scale linearly with the number of sites and can in principle deal with any number of atoms. One disadvantage of the positive-P representation is that the integration can show a tendency to diverge at short times for high collisional nonlinearities~\cite{Steel}. As long as the procedures followed to derive the Fokker-Planck equation for the positive-P function are valid~\cite{SMCrispin}, the stochastic solutions are guaranteed to be accurate wherever the integration converges. With all the results shown here, the solutions were found without any signs of divergences.

The system is very simple, with three potential wells in a linear configuration. Each of these can contain a single atomic mode, which we will treat as being in the lowest energy level. Atoms in each of the wells can tunnel into the nearest neighbour potential, with tunnelling between wells $1$ and $2$, and $2$ and $3$. With all the population initially in the middle well, the system acts as a periodic beamsplitter and mode recombiner.  
With the $\hat{a}_{j}$ as bosonic annihilation operators for atoms in mode $j$, $J$ representing the coupling between the wells, and $\chi$ as the collisional nonlinearity, we may now write our Hamiltonian.
Following the usual procedures~\cite{BHJoel}, we find
\begin{equation}
{\cal H} =  \hbar\sum_{j=1}^{3}\chi \hat{a}_{j}^{\dag\;2}\hat{a}_{j}^{2}
-\hbar J\left(\hat{a}_{1}^{\dag}\hat{a}_{2}+\hat{a}_{2}^{\dag}\hat{a}_{1}+\hat{a}_{3}^{\dag}\hat{a}_{2}
+\hat{a}_{2}^{\dag}\hat{a}_{3}\right).
\label{eq:Ham}
\end{equation}

To solve the full quantum equations, we use the positive-P representation~\cite{Pplus}, which allows for exact solutions of the dynamics arising from the Hamiltonian of Eq.~\ref{eq:Ham}, in the limit of the average of an infinite number of trajectories of the stochastic differential equations in a doubled phase-space. In practice we obviously cannot integrate an infinite number of trajectories, but have used numbers large enough that the sampling error is within the line thicknesses of our plotted results.
Following the standard methods~\cite{DFW}, the set of It\^o stochastic differential equations~\cite{SMCrispin} are found as
\begin{eqnarray}
\frac{d\alpha_{1}}{dt} &=& -2i\chi\alpha_{1}^{+}\alpha_{1}^{2}+iJ\alpha_{2}
+\sqrt{-2i\chi\alpha_{1}^{2}}\;\eta_{1},\nonumber\\
\frac{d\alpha_{1}^{+}}{dt} &=& 2i\chi\alpha_{1}^{+\,2}\alpha_{1}-iJ\alpha_{2}^{+}
+\sqrt{2i\chi\alpha_{1}^{+\;2}}\;\eta_{2},\nonumber\\
\frac{d\alpha_{2}}{dt} &=& -2i\chi\alpha_{2}^{+}\alpha_{2}^{2}+iJ\left(\alpha_{1}
+\alpha_{3}\right)
+\sqrt{-2i\chi\alpha_{2}^{2}}\;\eta_{3},\nonumber\\
\frac{d\alpha_{2}^{+}}{dt} &=& 2i\chi\alpha_{2}^{+\,2}\alpha_{2} - iJ\left(\alpha_{1}^{+}
+\alpha_{3}^{+}\right)
+\sqrt{2i\chi\alpha_{2}^{+\,2}}\;\eta_{4},\nonumber\\
\frac{d\alpha_{3}}{dt} &=& -2i\chi\alpha_{3}^{+}\alpha_{3}^{2}+iJ\alpha_{2}
+\sqrt{-2i\chi\alpha_{3}^{2}}\;\eta_{5},\nonumber\\
\frac{d\alpha_{3}^{+}}{dt} &=& 2i\chi\alpha_{3}^{+\,2}\alpha_{3} - iJ\alpha_{2}^{+}
+\sqrt{2i\chi\alpha_{3}^{+\;2}}\;\eta_{6},
\label{eq:Pplus}
\end{eqnarray}
where the $\eta_{j}$ are standard Gaussian noises with $\overline{\eta_{j}}=0$ and $\overline{\eta_{j}(t)\eta_{k}(t')}=\delta_{jk}\delta(t-t')$. As always, averages of the positive-P variables represent normally ordered operator moments, such that, for example, $\overline{\alpha_{j}^{m}\alpha_{k}^{+\,n}}\rightarrow\langle\hat{a}^{\dag\,n}\hat{a}^{m}\rangle$. We note that, while $\overline{\alpha_{j}}=\overline{(\alpha_{j}^{+})}^{\ast}$, $\alpha_{j}^{\ast}\neq\alpha_{j}^{+}$ on individual trajectories, and it is this freedom that allows classical variables to represent quantum operators.

\begin{figure}
\begin{center}
\includegraphics[width=0.8\columnwidth]{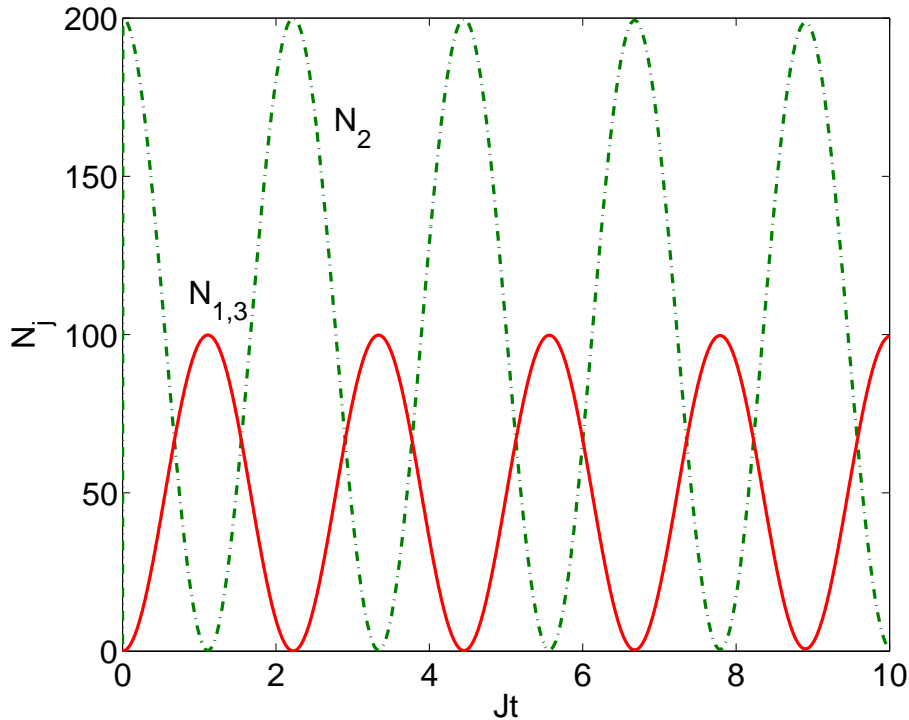}
\end{center}
\caption{(Color online) The populations in each well as a function of time, for $J=1$, $\chi=10^{-3}$, and $N_{2}(0)=200$, with $N_{1}(0)=N_{3}(0)=0$. The atoms in the centre well begin in a Fock state, although an initial coherent state leads to indistinguishable results. The results shown are the average of $1.08\times 10^{6}$ stochastic trajectories. The quantities plotted in this and subsequent plots are dimensionless.}
\label{fig:populations}
\end{figure}

As well as the populations in each well, we also calculate two types of quantum correlations. The first class of correlations are the number variances, including the number difference between the populations of wells $1$ and $3$. In the positive-P formulation, these are written as
\begin{eqnarray}
V(N_{j}) &=& \overline{ \alpha_{j}^{+\,2}\alpha_{j}^{2}}+\overline{\alpha_{j}^{+}\alpha_{j}}-\overline{\alpha_{j}^{+}\alpha_{j}}^{2},\nonumber\\
V(N_{1}-N_{3}) &=& V(N_{1})+V(N_{3})-2V(N1,N3),\nonumber\\
&=& V(N_{1})+V(N_{3})-2\left(\overline{\alpha_{1}^{\dag}\alpha_{1}\alpha_{3}^{\dag}\alpha_{3}} -\overline{\alpha_{1}^{+}\alpha_{1}}\times\overline{\alpha_{3}^{+}\alpha_{3}}\right)
\label{eq:variances} 
\end{eqnarray}
with these all giving values of zero for uncorrelated Fock states. Whenever one of the variances is less than the mean population of that mode, we have suppression of number fluctuations below the coherent state level. 

\begin{figure}
\begin{center}
\includegraphics[width=0.8\columnwidth]{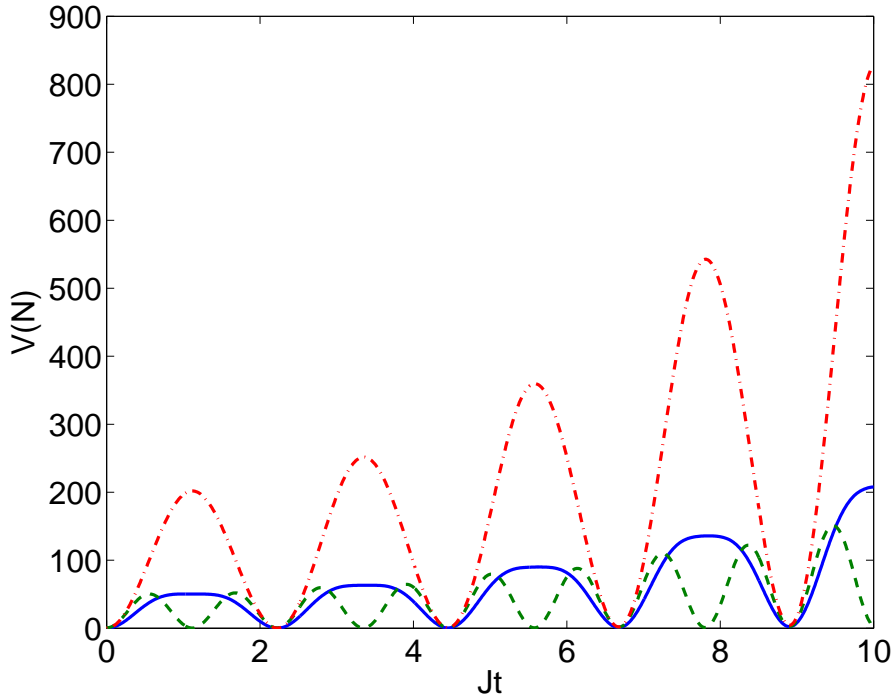}
\end{center}
\caption{(Color online) The number variances for the same parameters as Fig.~\ref{fig:populations}. The solid line is $V(N_{1})$, the dashed line is $V(N_{2})$, and the dash-dotted line is $V(N_{1}-N_{3})$, all averaged over $3.97\times 10^{5}$ trajectories. We see that the variances are periodic and that the maximum variances increase with time, becoming larger as the entanglement signature disappears.}
\label{fig:VN}
\end{figure}

The second correlation is an entanglement measure adapted from an inequality developed by Hillery and Zubairy, who showed that, considering two separable modes denoted byß $i$ and $j$~\cite{HZ},
\begin{equation}
| \langle \hat{a}_{i}^{\dag}\hat{a}_{j}\rangle |^{2} \leq \langle \hat{a}_{i}^{\dag}\hat{a}_{i}\hat{a}_{j}^{\dag}\hat{a}_{j}\rangle,
\label{eq:HZ}
\end{equation}
with the equality holding for coherent states. The violation of this inequality is thus an indication of the inseparability of, and entanglement between, the two modes. Cavalcanti \etal\cite{ericsteer} have extended this inequality to provide indicators of Einstein-Podolsky-Rosen (EPR) 
steering~\cite{Einstein,Erwin,Wiseman} and Bell violations~\cite{Bell}. We now define the correlation function
\begin{equation}
\xi_{13} = \langle \hat{a}_{1}^{\dag}\hat{a}_{3}\rangle\langle \hat{a}_{1}\hat{a}_{3}^{\dag}\rangle - \langle \hat{a}_{1}^{\dag}\hat{a}_{1}\hat{a}_{3}^{\dag}\hat{a}_{3}\rangle,
\label{eq:xi13}
\end{equation}
for which a positive value reveals entanglement between modes $1$ and $3$. We easily see that $\xi_{13}$ gives a value of zero for two independent coherent states and a negative result for two independent Fock states. This inequality, and the EPR steering development of it, have been shown to detect both inseparability and asymmetric steering in a three-well Bose-Hubbard model under the process of coherent transfer of atomic population 
(CTAP)~\cite{myJPB,myJOSAB}. 

\begin{figure}
\begin{center}
\includegraphics[width=0.8\columnwidth]{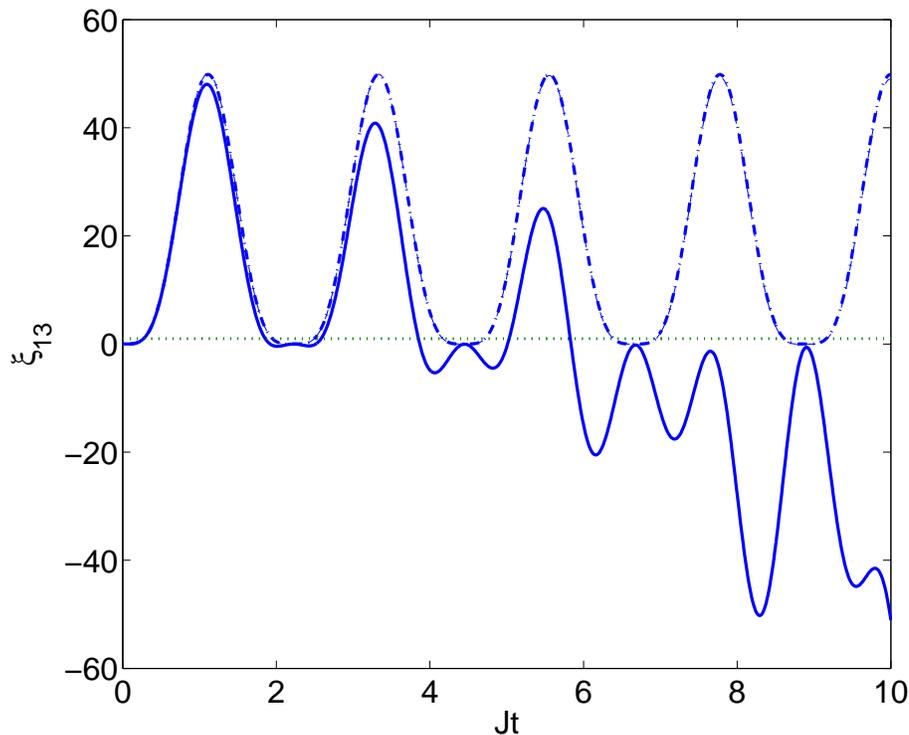}
\end{center}
\caption{(Color online) The entanglement criteria, $\xi_{13}$, as a function of time, for $J=1$ and an initial Fock state in the middle well. The solid line is for $\chi=10^{-3}$ ($1.08\times 10^{6}$ trajectories), while the almost indistinguishable dashed and dashed-dotted lines are for $\chi=10^{-4}$ ($3.64\times 10^{5}$ trajectories), and $\chi=10^{-5}$ ($1.36\times 10^{6}$ trajectories). We see that the entanglement is more persistent for lower non-linearities. The line at zero is a guide to the eye.}
\label{fig:xi13Fock}
\end{figure}


In all the results presented here, we begin with $200$ atoms in the middle well, with the other two being empty. We begin with these atoms initially in either Fock or coherent states, modelled as in ref.~\cite{states}. We note that this allows us to sample the appropriate positive-P distributions for these states using the Gaussian random number generator found in Matlab, for example. The equations were numerically integrated over a large number of stochastic trajectories, which was different for each result shown. This is an artefact of the program we use, which runs until we stop it, and then takes the averages. The times that we leave it running are not identical, therefore the numbers of trajectories are different, but this is not important as long as there are sufficient trajectories so that the sampling error becomes insignificant. We found that an initial coherent state gave the same average populations in each well as for a Fock state, but no entanglement was found, according to the measure of Eq.~\ref{eq:xi13}, and independently of the nonlinearity used.

The populations of each well as a function of scaled time $Jt$ are shown in Fig.~\ref{fig:populations}. As expected, we see that the average populations of wells $1$ and $3$ are identical. We also see that the oscillations are highly regular over the time investigated, with no sign of the damping of oscillations seen in other Bose-Hubbard systems~\cite{BHJoel,Chiancathermal}. While this may happen for higher collisional nonlinearities, we did not investigate this because they were seen to degrade the entanglement. On this scale, the results for $\chi=0$ are indistinguishable from those for $\chi=10^{-5}$. In Fig.~\ref{fig:VN} we show the number variances for the quantities $N_{1}\:(=\hat{a}_{1}^{\dag}\hat{a}_{1})$, $N_{2}$, and $N_{1}-N_{3}$, the population difference between the two initially empty wells. As the tunnelling Hamiltonian is equivalent to that of a beamsplitter, we do not expect the interaction to produce any squeezing such as would be expected from pair production. Consistent with this, we see that all the variances, while oscillatory, evolve under an envelope which increases with time. The number variance of a coherent state is equal to the mean number, and the individual variances for $N_{1}$ and $N_{2}$ stay below this level for some time, while the variance in the population difference $N_{1}-N_{3}$ has risen above this level by the second oscillation. This shows that the tunnelling adds noise to the system, which is to be expected because the tunnelling in each direction is independent. At the level of individual particles, the tunnelling in each direction is random, so that any initial sub-Poissonian statistics will evolve toward being Poissonian.

We now turn to the calculation of $\xi_{13}$ of Eq.~\ref{eq:xi13}, our chosen entanglement witness. For initial coherent states we found no evidence of entanglement at all, independent of the strength of the $\chi$ nonlinearity. This is consistent with our previous result for coherent population transfer~\cite{myJOSAB}, where entanglement was also not found for an initial coherent state. When we consider an initial Fock state of fixed number, we do find evidence of entanglement, as shown in Fig.~\ref{fig:xi13Fock}. We investigated three different positive values of $\chi$, finding that the stronger interactions tend to degrade the predicted entanglement as time increases. For nonlinearites $\chi=10^{-4}$ and $10^{-5}$ we see that the signature of the entanglement is periodic, with no sign of degradation up to $Jt=10.$ On the other hand, with $\chi=10^{-3}$, the function stays negative after three oscillations, with the first peak being noticeably higher than the other two. We also calculated the EPR steering extension of $\xi_{13}$, but found no positive values for any times or nonlinearities, indicating that the degree of inseparability produced is not sufficient for a demonstration of this property.

The nonlinear interaction, known from previous work to produce squeezing~\cite{Simonsqueeze} and non-Gaussian entanglement and EPR 
steering~\cite{NGJoel}, actually degrades the chosen quantum correlations in this system. In the nonlinear coupler~\cite{nlc}, comprised of two evanescently coupled Kerr waveguides in an optical cavity, the nonlinearity creates the necessary quantum states since the cavity is pumped with a coherent input. The difference from the present system is that the nonlinearity is smaller and both interactions have many cavity lifetimes to create the quantum correlations. In quantum optics, as illustrated in Ref.~\cite{NGJoel}, one simple method of obtaining entanglement is to put a coherent state through a nonlinear interaction so as to produce a quantum state of the electromagnetic field, such as a squeezed state. This can then be mixed on a beamsplitter, either with vacuum or another quantum state, with the outputs being entangled~\cite{bstangle}. That this does not happen with our model when starting from a coherent state here indicates that the collisional nonlinearity does not have time to form a sufficiently quantum state before the tunnelling takes effect. 

\begin{figure}
\begin{center}
\includegraphics[width=0.8\columnwidth]{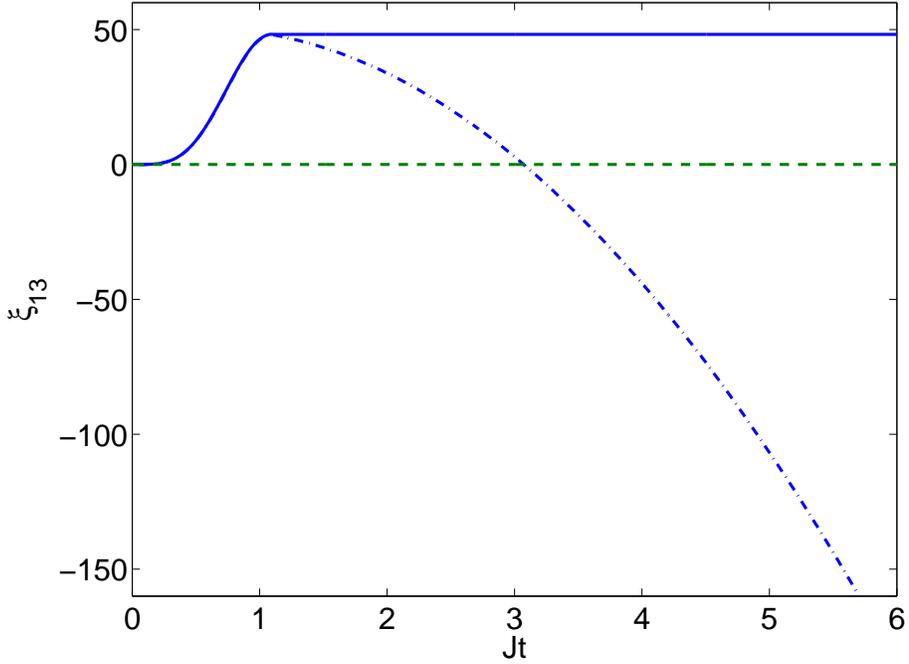}
\end{center}
\caption{(Color online) The entanglement criteria, $\xi_{13}$, as a function of time, for pulsed tunneling, with $J=1$ turned off at $t=\tau_{p}\:(\approx 1.1)$, the first time of maximum population transfer. The solid line is for $\chi=10^{-3}$ also turned off at time $\tau_{p}$ ($7.35\times 10^{5}$ trajectories), while the dashed-dotted line represents the evolution with 
$\chi$ constant ($9.1\times 10^{5}$ trajectories). The line at zero is a guide to the eye. We see that the nonlinearity prevents this measure from registering entanglement after approximately one more of the oscillatory periods shown in Fig.~\ref{fig:populations}. }
\label{fig:Fpulse}
\end{figure}

Two methods suggest themselves to surmount this difficulty with our system. We can either hold the middle well isolated until the nonlinearity has acted sufficiently to form an appropriate quantum state, or turn the tunnelling off at the first maximum of population transfer. Because we do not know a priori how long a sufficiently squeezed state will take to develop, we do not investigate this option here. We will instead use the freedom in engineering optical potentials that is being developed at the present time~\cite{Mark} and assume that the two end wells can be changed and moved at the time of the first maximum of population transfer, which is also the time of the maximal entanglement signature. Labelling this time as $\tau_{p}$, we use a time dependent $J(t)=1-\Theta(\tau_{p})$, where $\Theta$ is the Heaviside step function. The results are shown in Fig.~\ref{fig:Fpulse} as the dashed line, where we see that the entanglement signature begins to decay as soon as the tunnelling is turned off. This suggests turning the nonlinear interaction off at the same time as the tunnelling, possible in principle via Feshbach resonance techniques, with the result of this shown as the solid line. Without the scattering term, the value of $\xi_{13}$ remains constant as we essentially have the free evolution of harmonic oscillators. Although our analysis here is not designed to model an actual experiment with all the attendant noise sources, it does point to a possible method for the achievement of spatially isolated entangled atomic samples.


In conclusion, we have proposed and analysed a simple atomic entangling analogue of a combined beamsplitter and mode recombiner, showing that it can be used to manufacture spatially separated entangled atomic modes. This will be a step towards removing the locality loophole from tests of entanglement and EPR steering for massive bosons. We have performed a fully quantum analysis of our model, making no approximations in the actual calculations. Using an inequality suited to systems with number conservation, we have calculated the entanglement available, and shown how to preserve this by turning off the tunnelling and collisional interactions. We have also shown that, in order to produce entangled modes by this method, having an initial state with quantum correlations is more important than the interactions. In fact, given an initial non-classical state, the collisional nonlinearity only acts to degrade the performance. As a Fock state is the natural state of a single atomic mode of an isolated well, it is to our advantage that this state can be used to produce a clear entanglement signal. We realise there are problems with the reproduction of such a state for a series of experimental runs, and will be addressing these in a subsequent work. The conceptual simplicity of our system suggests that there should be no insurmountable barriers to an experimental realisation. 

{\it Acknowledgments} This research was supported by the Australian Research Council under the Future Fellowships Program (Grant ID: FT100100515). We acknowledge fruitful discussions with Joel Corney, Simon Haine, and Mark Baker.

\end{document}